\documentclass[12pt]{article}

\def\be{\begin{equation}}
\def\ee{\end{equation}}
\def\bea{\begin{eqnarray}}
\def\eea{\end{eqnarray}}
\def\vp{\varphi}
\def\K{K{\"a}hler}
\newcommand{\rf}[1]{(\ref{#1})}

\newcommand{\mr}{\mathcal{R}}

\usepackage{epsfig}

\usepackage{amssymb,latexsym}
\usepackage{graphicx}
\usepackage{amssymb, amsmath, amsopn, amsthm}
\usepackage[section]{placeins}

\makeatletter \@addtoreset{equation}{section}
\makeatother

\begin{document}
\thispagestyle{empty}
\hskip 1 cm
\vskip 0.5cm

\vspace{25pt}
\begin{center}
    { \LARGE{\bf  Planck 2013 and Superconformal Symmetry}\footnote{
Based on the lecture at the Les Houches School ``Post-Planck Cosmology,'' 2013}}
    \vspace{33pt}

  {\large  {\bf  Renata Kallosh}}

    \vspace{10pt}

    \vspace{10pt} {Department of Physics,
    Stanford University, Stanford, CA 94305}

    \vspace{20pt}
 \end{center}

\begin{abstract}
We explain why the concept of a spontaneously broken superconformal symmetry is useful to describe inflationary models favored by  Planck. Non-minimal coupling of complex scalars to curvature, ${\cal N}( X,\bar X) \, R$,  is compulsory for superconformal symmetry. Here  ${\cal N}$   is the \K\,  potential of the embedding moduli space, including the inflaton and the conformon.   It appears that  such a non-minimal coupling  allows  generic chaotic models of inflation to reach an agreement with the observable  $(n_{s},r)$ values. 

We describe here the superconformal versions of the cosmological attractors whose bosonic part was presented in  lectures of 
A. Linde in this volume.  A distinguishing feature of this class of models is that they tend to lead to very similar predictions which are not very sensitive with respect to strong modifications of the theory. The (super)conformal  symmetry underlying (super)gravity allows a universal description of a large class of models which agree with observations and predict the tensor to scalar ratio $10^{-3} \lesssim r \lesssim 10^{-1}$.

\

\end{abstract}

\newpage

\tableofcontents

\parskip 8.5pt

\section{Introduction}

The recent data by Planck \cite{Ade:2013uln}, \cite{Spergel:2013rxa} and the earlier results by WMAP \cite{Bennett:2012zja}, as well as the combined results by WMAP,  ACT and SPT \cite{Calabrese:2013jyk},  support the models of a single-field inflation, and strongly constrain these models.  Successful models have to predict significant red tilt of the scalar curvature perturbations with the spectral index $n_s = 0.960 \pm 0.007$ \cite{Ade:2013uln} \footnote{It is about one $\sigma$ higher in \cite{Spergel:2013rxa}  and given by $n_s = 0.967 \pm 0.007$, and also a bit higher in other pre-Planck CMB measurements as discussed in \cite{Calabrese:2013jyk}.}, and a tensor-to-scalar ratio, $r < 0.15
$. In view of such simplicity (and absence of non-gaussianity) one can ask a question: Is there any symmetry behind the successful versions of chaotic  inflation \cite{Linde:1983gd}? Spontaneously broken superconformal symmetry 
is one possible answer.

For many years, nonminimal coupling to gravity ${1\over 2} \Omega(\phi) R$ was considered as something exotic, at least in the context of inflation. It was of course known that the non-minimal coupling to gravity is required for the Weyl invariance of the action of scalars interacting with gravity. However,  conformal invariance implied traceless energy-momentum tensor, and therefore the energy of the scalar field was decreasing very fast, as $a^{-4}$, i.e. in the same way as in the hot Big Bang. It was possible to introduce inflationary theories with broken local conformal invariance, but typically this symmetry breaking was explicit rather than spontaneous. In other words, the full theory was not invariant under the Weyl transformations.

Meanwhile the standard formulation of supergravity is based on superconformal symmetry, which becomes {\it spontaneously broken} when certain fields or their combinations acquire non-zero vacuum expectation values. In these cases, the superconformal symmetry becomes well hidden, but it is still present even after it is spontaneously broken. This resembles the situation in the standard model of electroweak interactions, which remains gauge-invariant even after the Higgs field acquires its vev.

Thus, in the context of supergravity, unlike in the ordinary GR, the concept of (super)conformal invariance is quite natural.
It serves as a basic principle which helps to formulate supergravity in a consistent way. 

Back in  2000, it was proposed to use the superconformal symmetry underlying supergravity in application to cosmology \cite{Kallosh:2000ve}. Later on, in 2010, the construction of a consistent supersymmetric version \cite{Ferrara:2010yw},\cite{Einhorn:2009bh}  of the Higgs inflation 
\cite{Futamase:1987ua},   \cite{Sha-1}
   was based on a superconformal version of the theory.
The data from  Planck in 2013 support the concept of the superconformal symmetry since it allows to explain why many models enhanced by the non-minimal coupling to gravity tend to agree with observations.

Models with non-minimal coupling of scalars to gravity of the type $\sqrt{-g} \, {1\over 2} \Omega(\phi) \, R(g) $ are associated with the so-called Jordan frame. One can make a Weyl transformation of the metric to the Einstein frame \be
\sqrt{-g} \, {1\over 2} \Omega(\phi) \, R(g) \qquad \Rightarrow  \qquad g_{\mu\nu}'= \Omega(\phi) g_{\mu\nu}  \qquad \Rightarrow  \qquad \sqrt{-g'}\, {1\over 2} R(g') \ ,
\ee
so that the curvature term is decoupled from scalars, $\sqrt{-g'}\,  {1\over 2} R(g')$, and work with an equivalent model with a minimal coupling to gravity. 
However,  the original Jordan frame with non-minimal coupling to gravity has significant advantages, like  symmetries, simple potentials and other features which lead to the concept of the cosmological attractors in the $(n_s, r)$ plane. One can reproduce all cosmological results also in the Einstein frame, in standard supergravity, however,  the superconformal symmetry provides the cosmological attractors with a natural starting point  and explains the  universality feature.

Here we present the superconformal versions of the  cosmological inflationary attractors whose bosonic part was described in the lectures by 
A. Linde in this volume \cite{A}.  We explain the technical aspects of the N=1 superconformal symmetry which play an important role for cosmology.
We hope this lecture will be useful for cosmologists interested in inflation, who are familiar with standard supergravity, as well as for supersymmetry experts familiar with black hole attractors. There is an opportunity to use the experience with the supersymmetric attractor behavior of the moduli near black hole horizon in application to cosmology.

\

\section{Superconformal theory underlying supergravity}

The most familiar form of supergravity which was frequently used in cosmological applications is based on the so-called chiral matter multiplets, whose first component is a complex scalar field. The theory is codified by a \K\, potential $K(z, \bar z)$ and a superpotential $W(z)$. The scalar-gravity part of the supergravity action {\it in the Einstein frame describes the minimal coupling of the curvature $R$ to scalars}, with the following kinetic term and the potential
for $n$ complex scalars $z^i, i=1,...,n$ are (in units $M_{Pl}=1$)
\be
{1\over \sqrt{-g}} {\cal L}_{\rm sg}=  {1\over 2} R + g_{i, \bar j}  \partial_\mu z^i\partial^\mu \bar z^{\bar j}  -  e^{K}(|DW|^2- 3 |W|^2)
\, , \qquad g_{i, \bar j}\equiv {\partial^2  K(z, \bar z)\over \partial z^i \partial \bar z^{\bar j} }\ .\label{standard}\ee
Here the metric of the `physical' moduli space, $g_{i, \bar j}$,  is positive definite.

This form of supergravity, called Poincar\'e supergravity  is obtained from the underlying superconformal theory which in addition to local supersymmetry of a Poincar\'e supergravity has also extra local symmetries: Weyl symmetry,  $U(1)\, \mr$ symmetry, special conformal symmetry and special supersymmetry. To have these additional symmetries one has to have an extra chiral superfield, a so-called conformal compensator\footnote{These were first introduced in  \cite{Siegel:1977hn}
\cite{Kaku:1978ea}. Superconformal symmetry underlying supergravity  in general case was developed  in 
\cite{Cremmer:1978hn}. The textbook  \cite{Freedman:2012zz} provides a full description of N=1 superconformal models as well as the derivation of a standard supergravity \rf{standard} from the superconformal models \rf{sc}.}
so that the model depends on $n+1$ chiral superfields $X^I, \bar X^{\bar J}, I=0,1,...,n$. 

To fully specify the superconformal theory one has to give an information on ${\cal N} (X,\bar X)$ and on ${\cal W}(X)$ (and on the vector multiplets coupling $f_{AB}$).
The complete action with all other fields present, vectors and fermions, can be found in \cite{Freedman:2012zz}.

Here we are interested in the scalar-gravity part of the superconformal $SU(2,2|1)$ invariant action 
\be
{1\over \sqrt{-g}}\mathcal{L}_{\rm sc}=-\frac 1 6{\cal N} (X,\bar X)R
-G_{I\bar J}{\cal D}^\mu X^I\,{\cal D}_\mu \bar X^{\bar J}-G^{I\bar J}{\cal W}_I \bar{{\cal W}}_{\bar J} \, , \qquad I, \bar I = 0,1,...,n.
 \label{sc}
\end{equation}
Here, ${\cal N} (X,\bar X)$ is the \K\, manifold of the embedding space with coordinates $X^I, \bar X^{\bar J}, I=0,1,...,n$. 
The $U(1)\, \mr$ symmetry covariant derivative   is
$
{\cal D}_\mu X^I = \partial_\mu X^I - i A_\mu
X^I   $,
where   $A_\mu$ is a gauge field of the 
the  local $U(1)\, \mr$ symmetry.
The metric of the embedding space is
\be
  G_{I\bar J}=\partial_I \partial_{\bar J} {\cal N} \equiv {\partial  \mathcal{N}(X, \bar X)\over \partial X^I \partial \bar X^{\bar J}} \ .
\ee
It is not positive definite, the signature is $-, +,..., +$. One negative direction in the metric of the embedding space signals the presence of the conformal compensator, the field which may be removed from the theory, using the Weyl symmetry.
The F-term potential 
$
V_F= G_{I\bar J}  F^I \bar F^{\bar J}=G^{I\bar J}{\cal W}_I \bar{ {\cal W}}_{\bar J} 
$
is constructed from the superpotential ${\cal W}$ and 
$
{\cal W}_I\equiv {\partial {\cal W}\over \partial X^I}$,   $\bar{{\cal W}}_{\bar J}\equiv  {\partial \bar{{\cal W}}\over \partial \bar X^{\bar J}}
$.
Here  $F^I$ is an auxiliary field whose value is defined by the derivatives of the superpotential $F^I= G^{I\bar J} \bar {\cal W}_{\bar J}$.
In the absence of fermions  the action (\ref{sc}) has a  local conformal and local $U(1)$ ${\mathcal{R}}$-symmetry, which is part of the superconformal $SU(2,2|1)$ 
symmetry. This means that the action (\ref{sc}) is invariant under the following transformations:
\be
(X^I)'= e^{\sigma(x)+\Lambda(x)}  X^I\, , \qquad (\bar X^I)'= e^{\sigma(x)-\Lambda(x)}  \bar X^I  \ ,
\ee
\be
g_{\mu\nu} '= e^{-2\sigma(x)} g_{\mu\nu}\, , \qquad A_\mu '= A_\mu +\partial _\mu \Lambda(x)
 \ .
\label{dil}\ee
By construction, the action requires   a non-minimal coupling of scalar to the curvature $R$. Since the action has a local conformal invariance under which $R$ has a conformal weight $w=2$, the function of scalars to which it is coupled, ${\cal N} (X,\bar X)$,  must have a conformal weight $w=2$, so that the action is Weyl invariant. In  d=4 $\sqrt{-g}$ has   conformal weight $w=-4$ and therefore
$\sqrt{-g} \, {\cal N} (X,\bar X) \, R$ has  conformal weight $w=0$. Thus, the model with minimal coupling of curvature to scalars, with ${\cal N} (X,\bar X)=const$, cannot describe a Weyl invariant theory. {\it Therefore a function of scalars in front of $R$ is compulsory in theories with superconformal symmetry}.

We would like to stress here that the local conformal symmetry of the action \rf{sc} does not require the scalar coupling to be of the form $-{1\over 12} \phi^2 R$, which is familiar case in general relativity with the canonical kinetic term for a scalar, $-{1\over 2} (\partial_\mu \phi)^2$. In \rf{sc} the kinetic terms for scalars are defined by a \K\, geometry which allows to preserve the local conformal symmetry for more general couplings. A detailed explanation of this phenomena is given in section  on `Bosonic Conformal $\Delta$-Model: Simplified Superconformal Model' in \cite{Kallosh:2013pby}.

\subsection{Canonical Superconformal Supergravity (CSS) \cite{Ferrara:2010yw}}
Consider the class of models useful for cosmology. Various  deformation of   `Canonical Superconformal Supergravity ' models introduced  in \cite{Ferrara:2010yw} lead to cosmological attractors. The  chiral multiplets in our model $X^I$, include the compensator field $X^0$, the inflaton  $X^1=\Phi$ and the Goldstino  superfield $X^2=S$
\be
X^I= (X^0, X^1=\Phi, X^2=S) \ .
\ee

 The role of the conformon field $X^0$  in the action is to support an unbroken  superconformal symmetry. During inflation the conformon breaks spontaneously the Weyl symmetry and the $\mathcal{R}$ $ U(1)$ symmetry when equations of motion of the superconformal theory  are solved, for example,  with an ansatz
$X^0=\bar X^{\bar 0}= \sqrt 3 M_P$.  In all models of interest a special effort is taken to provide $S=0$ to be a minimum of the potential during inflation, i. e. the scalar partner of a fermion called godstino, a sgoldstino,  must vanish during inflation.

The \K\, potential in CSS is flat and has an $SU(1,2)$ symmetry
\be
{\cal N} (X,\bar X)= -|X^0|^2 +|\Phi|^2+ |S|^2 \ .
\label{can}\ee
 This means that  
\be
G_{I\bar J}={\cal N}_{I\bar J}= \eta_{I\bar J} \,, G^{I\bar J}= \eta^{I\bar J}, \qquad I=0, 1,2 \ .
\ee
 , and
$
 \eta_{0\bar 0}=-1,  \, \eta_{\Phi \bar \Phi }=\eta_{S \bar S }=1 \,.
 $
The superpotential is conformon-independent
\be
{\cal W}_0\equiv {\partial {\cal W} \over \partial X^0}=0 \ .
\ee
For example,  we choose a cubic  $X^0$-independent superpotential: 
 \be
{\cal W}(\Phi, S)
=   \sqrt \lambda \, \, S\,  \Phi^2\,  \ .
\label{superpot}\ee
The superconformal action is simple, the scalar-gravity part is
 \begin{eqnarray}
\sqrt{-g} \Big [{1\over 6}(|X^0|^2 - |\Phi|^2 - |S|^2) R
+D^\mu X^0\,D_\mu \bar X^{\bar 0}-D^\mu \Phi\,D_\mu \bar \Phi -D^\mu S\,D_\mu \bar S
-\lambda (  |\Phi\bar \Phi|^2 + 4 |S\Phi|^2)\Big ].\nonumber\\
 \label{can1}
\end{eqnarray}
{\it Kinetic terms of scalars in CSS in Jordan frame are canonical since $\cal N$ is flat,  the potential is the same as in  globally supersymmetric theories }
\be
V=G^{I\bar J}{\cal W}_I \bar{{\cal W}}_{\bar J},   \quad I, \bar I = 0,1,2  \qquad \Rightarrow   \qquad V= \delta ^{I\bar J}{\cal W}_I \bar{{\cal W}}_{\bar J} ,   \qquad I, \bar I = 1,2\, .
\ee
 It is the simplicity of the CSS action which to some extent is preserved in deformations, which leads to a certain universality of the cosmological models.

\subsection{The role of sgoldstino in models with ${\cal W}= S f(X^0, X^1)$}
  The inflaton and sgoldstino together break spontaneously the local supersymmetry when the solution of equations of motion at $S=0$ has a non-vanishing auxiliary field $F_{S}(\Phi)\equiv {\partial {\cal W}\over \partial {S}}\neq 0$.  The potential during inflation depends on the inflaton as 
\be
V(\Phi)= |F_{S}(\Phi)|^2= \Big |{\partial {\cal W}(S, \Phi)\over \partial {S}}\Big |^2
\label{JordanV}\ee 
{\it under condition that the superpotential is linear in $S$ and during inflation $S=0$}. Note that the total potential of the superconformal theory in the Jordan frame is extremely simple, as eq.  \rf{JordanV} shows.

We will find that during inflation with $S=0$ the $S$ direction is indeed a sgoldstino direction since only in $S$ direction the auxiliary field $F_S$ is not vanishing,
\be
F_0= {\partial {\cal W}\over \partial X^0}=0\, ,\qquad F_\Phi= {\partial {\cal W}(S, \Phi)\over \partial \Phi}=0
\, ,\qquad F_S= {\partial {\cal W}(S, \Phi) \over \partial S} \neq 0 \ .
\ee
The reason for this is that in our class of models the scalar field $S$ vanishes during inflation and $S=0$ is a minimum of the potential.  More details on this can be found  in \cite{Kallosh:2010ug}.

\subsection{Spontaneous breaking of the Weyl and $U(1)\, \mr$ symmetries}
Using two gauge symmetries in \rf{dil} , a local Weyl symmetry with the parameter $\sigma(x)$ and the $U(1)\, \mr$ symmetry
with the parameter $\Lambda(x)$, one can either use the an algebraic type gauge-fixing procedure which does not require a propagating ghosts action for unitarity or,  alternatively,  look for the solutions of equation of motion of the theory which spontaneously break these gauge symmetries. In both cases, this allows to use two  conditions on some combinations of fields and insert these two conditions in the action. For example, any algebraic condition on a function of the scalar variables $X, \bar X$ will satisfy the condition: when viewed as a gauge-fixing condition in the path integral, there is no need for ghosts since $X, \bar X$ transform under Weyl and $U(1)\, \mr$ symmetries without derivatives, as shown in \rf{dil}. The same two functions of $X, \bar X$ imposed on the vev's of these scalars describe the solutions with spontaneously broken symmetry. Examples of such two functions which are  conditions of spontaneously broken symmetry, are presented in \cite{Kallosh:2000ve},\cite{Ferrara:2010yw} and in \cite{Freedman:2012zz}.

1. Spontaneous breaking of Weyl  symmetry leading to Einstein frame supergravity
\be
{\cal N} (X,\bar X)=-3\, , \qquad  - \frac 1 6{\cal N} (X,\bar X)R \Rightarrow {1\over 2} R \ .
\label{Einst}\ee
There are choices of  $U(1)\, \mr$ symmetry breaking, for example one can impose that $\bar X^0= X^0$.
Together with \rf{Einst} these two conditions allow to replace $n+1$ complex coordinates of the embedding space $X^I, \bar X^{\bar J}, I=0,1,...,n$ by $n$ complex physical supergravity scalars $z^i, \bar z^{\bar i}, i= 1,...,n$ and lead to an action \rf{standard}.

2. Spontaneous breaking of Weyl  symmetry leading to an arbitrary Jordan frame supergravity.

These  conditions were introduced in the context of a supersymmetric Higgs inflation and studied in general case in \cite{Ferrara:2010yw}
\be
{\cal N} (X,\bar X)= -3 \, \Omega (z, \bar z)\, , \qquad  - \frac 1 6{\cal N} (X,\bar X)R \Rightarrow  \frac 1 2 \Omega (z, \bar z)\, R \ .
\label{Jordan} \ee
Here, with account of the condition \rf{Jordan} supplemented with one condition  of the $U(1)\, \mr$ symmetry breaking one can use only physical scalars $z^i, \bar z^{\bar i}, i= 1,...,n$, which leads to the supergravity action in the Jordan frame presented in \cite{Ferrara:2010yw}. The relation between the non-minimal coupling between physical fields $z, \bar z$ of supergravity and the \K\, potential $K(z, \bar z)$ of standard supergravity is the following, 
\be
 \Omega (z, \bar z)= e^{-{1\over 3} K(z, \bar z)} \ .
\label{kahler}\ee

\section{Deformation of CSS}
There are  few classes of deformation of  CSS which we have found useful for cosmological applications. 

1.  The first class corresponds to a critical point $\Delta=\pm {1\over 6} $, where $\Delta$ is a parameter of a particular deformation of $\mathcal{N}\left( X,\bar{X}\right)$,   from a quadratic expression \rf{can}. The \K\, potential of the embedding space depending on $X^0$ and $X^1= \Phi$ is deformed, it is not  flat anymore (we will discuss the dependence on the goldstino multiplet $S$ separately). 
\begin{equation}
\mathcal{N}\left( X,\bar{X}\right) =-\left| X^{0}\right| ^{2}+\left|
X^{1 }\right| ^{2}- 3\, \Delta  |X^0|^2  \left[ \left(\frac{X^{1}}{X^{0}}\right)^{2}+\left(\frac{\bar X^{\bar 1}
}{\bar X^{\bar 0}}\right)^{2}\right]   .
\label{calN}\end{equation}
 In these models the scalars are coupled to supergravity as ${\xi\over 2}  \phi^2 R$, where $\Delta = \pm  \Big ({1\over 6}+\xi \Big )$. A more  complicated coupling is possible when additional non-canonical terms are present in $\mathcal{N}\left( X,\bar{X}\right)$.
The superpotential may or may not be deformed from the case of the conformon-independent case as in CSS.

2. The second  class corresponds to a critical point $\Delta=0$. The \K\, potential of the embedding space $\mathcal{N}\left( X,\bar{X}\right)$ remains flat (with exception of the quartic in sgoldstino terms, required for stabilization) but the superpotential requires the dependence on the conformon field $X^0$. In these models the scalars are conformally coupled to gravity as $\pm {1\over 6}  \phi^2 R$. The superpotential is  deformed from CSS in various ways and it depends on the conformon.

3. More general deformation of the canonical superconformal supergravity models may also be studied, as we will show in the case of $\alpha$-attractors.

We will describe the superconformal origin of various recent cosmological attractor models presented in lectures by Linde \cite{A} in this volume. They will be classified according to the three classes of deviation from CSS given above.

\subsection{Deviation from the critical point $\Delta_{cr}=\pm {1\over 6} $, \cite{Kallosh:2013pby}}
If  we replace $\Delta $ by $\pm  ( \xi+1/6) $, the \K\, potential of the embedding space \rf{calN} becomes
\begin{equation}
\mathcal{N}\left( X,\bar{X}\right) =-\left| X^{0}\right| ^{2}+\left|
X^{1 }\right| ^{2}\mp  3 \Big ( \xi+{1\over 6}\Big )  |X^0|^2  \left[ \left(\frac{X^{1}}{X^{0}}\right)^{2}+\left(\frac{\bar X^{\bar 1}
}{\bar X^{\bar 0}}\right)^{2}\right]  .
\label{N1}\end{equation}
At $\xi=-1/6$ ($\Delta=0$)  the \K\, potential has an enhanced $SU(1,1)$ symmetry. 
\begin{equation}
\mathcal{N}\left( X,\bar{X}\right) =-\left| X^{0}\right| ^{2}+\left|
X^{1 }\right| ^{2}.
\label{N11}\end{equation}
The second symmetry becomes manifest  at $\xi=0$. Consider for simplicity only the case  $\Delta = \Delta_{\rm cr}=  1/6$. 
\begin{equation}
\mathcal{N}\left( X,\bar{X}\right) =
-\left | X^{0}\right| ^{2} + {1\over 2}  \left | X^{0}\right| ^{2}\left({X^1\over X^0} + {\bar X^{\bar 1}\over \bar X^{\bar 0}} \right)^2 .
\label{N2}\end{equation}
It is  invariant under the transformation with real $\Lambda$,
\bea\label{sym2}
X^1 \rightarrow  X^1 +  \Lambda X^0\, , \qquad 
\bar X^1 \rightarrow  \bar X^1 +  \Lambda \bar X^0\ , \qquad \Lambda=  \bar \Lambda\,  \ .
\eea
This transformation mixes the inflaton $X^1$ with the conformon $X^0$. In terms of the homogeneous coordinate $z= X^1/X^0$, for $\Delta= 1/6$ this symmetry is a shift symmetry  in the real  direction
\be
z\rightarrow z+\Lambda\, ,\qquad \bar z\rightarrow  \bar z + \Lambda \  , \qquad \Lambda=  \bar \Lambda\, .
\ee
For supergravity applications, this means that  for $\Delta= 1/6$ the \K\ potential does not depend on the real part of the field $\Phi$, which can be identified with the inflaton.

To summarize, the embedding \K\, potential  as a function of $\Delta=\pm  ( \xi+1/6)$ has two critical points with enhanced symmetry. One has  a maximal enhanced symmetry,   $\Delta=0,  \,   \xi_{\rm re}= -1/6, \,  \xi_{\rm im}= -1/6$ 
and the other is a   double critical point of enhanced symmetry,   $\Delta=\Delta_{\rm cr}= \pm 1/6, \,  \xi_{\rm re/im}= 0, \,   \xi_{\rm im/re}= -1/3$.
Here $ \xi_{\rm re}$ is the non-minimal coupling to gravity of the real part of  $X^1$ and  $ \xi_{\rm im}$ is the non-minimal coupling to gravity of the imaginary part of  $X^1$.

\subsubsection{Chaotic inflation in the theory with $V=\lambda \phi^4/4$  and with non-minimal coupling to gravity ${\xi\over 2} \phi^2 R$}
The superconformal action  for the model ${\xi\over 2} \phi^2 R- \lambda \phi^4/4$ in \cite{Kallosh:2013pby} is given by the expression in \rf{sc} with  $\mathcal{N}\left( X,\bar{X}\right)$ in \rf{calN}, which has to be supplemented by the dependence on $S$,  and  the superpotential in \rf{superpot}. But first  we consider, for simplicity, the non-supersymmetric locally conformal version of this model, which does not requires the sgoldstino, and where the potential is conformal, $V= \lambda (X^1 \bar X^{\bar 1} )^2$.

{\it Jordan frame}

\noindent We break Weyl symmetry spontaneously by imposing a condition
\be
X^0=\bar X^{\bar 0}= \sqrt 3 M_P\, .
\label{Pl}\ee
The action \rf{sc}  becomes 
\be
\sqrt{-g}^{-1}\mathcal{L}_{\rm sg}=   {1\over 2} M_P^2 R -  {1\over 6}  
X^{1 }\bar{X}^1  R\ +  {\Delta\over 2}   \left( (X^{1
})^2+(\bar{X}^{\bar{1 }})^2\right) R
-\partial^\mu X^1\,\partial_\mu \bar X^{\bar 1}-\lambda (X^1 \bar X^{\bar 1} )^2  \, .
 \label{SimplescGravJ}
\end{equation}
We see that the coupling $-  {1\over 6}  X^{1 }\bar{X}^1  R$ is still  conformal, but the holomorphic and anti-holomorphic couplings
${\Delta\over 2} \left( (X^{1
})^2+(\bar{X}^{\bar{1 }})^2\right) R$ will change coupling with gravity for the real and imaginary part of $X^1 \equiv {1\over \sqrt 2} \varphi= {1\over \sqrt 2} (\varphi_1 + i  \varphi_2)$:
\be
\sqrt{-g}^{-1}\mathcal{L}_{\rm sg}=    {1\over 2}  M_P^2 R -   {1\over 2}  \Big ({1\over 6} - \Delta \Big) \varphi_{1}^{2} R  -  {1\over 2} \Big({1\over 6}+ \Delta \Big)\varphi_{2}^{2}R 
-{ {1\over 2} } \partial^\mu \varphi_1\,\partial_\mu \varphi_1 - { {1\over 2} } \partial^\mu \varphi_2\,\partial_\mu \varphi_2-  {\lambda \over 4}  (\varphi_1^2 + \varphi_2^2)^2  \, .
 \label{SimplescGravJ1}
\end{equation}
The model has a symmetry
$
\Delta \rightarrow -\Delta\, , \, \varphi_1 \rightarrow \varphi_2 \, , \, \varphi_2 \rightarrow \varphi_1.
$
 We make a choice $\Delta = 1/6 +\xi>0$. If   the model has a minimum   at $ \varphi_2=0$  it  is reduced to
\be
\sqrt{-g}^{-1}\mathcal{L}_{\xi}=    {1\over 2}  M_P^2 R + {\xi\over 2}\, \varphi^2  \, R
-{1\over 2} \partial^\mu \varphi\,\partial_\mu \varphi -  {\lambda\over 4}  \varphi^4  \, ,
 \label{SimplescGravJ2}
\end{equation}
where $\varphi_1\equiv \varphi$. This is a $\varphi^4$ model with non-minimal coupling to gravity which agrees with Planck  at $\xi \gtrsim 0.002$. We will explain below how it is derived from the superconformal theory.

Thus, starting with spontaneously broken conformal symmetry, we have reproduced  the model ${\lambda\over 4} \phi^4 - {\xi\over 2}\phi^2 R$  as well as the Higgs inflation models \cite{Futamase:1987ua,Sha-1,Einhorn:2009bh,Ferrara:2010yw} where $\xi$ is a parameter of a non-minimal coupling to gravity. The difference is that in all these models, except \cite{Ferrara:2010yw}, local 
 conformal symmetry is absent (i.e. broken explicitly).  In the limit of large $\xi$ the bosonic model in Einstein frame approaches the Starobinsky model \cite{Starobinsky:1980te}, in the form depending on Einstein gravity interacting with the scalar field, which is dual to the original $R+R^2$ model.

 The new interpretation of the parameter $\xi$ in our superconformal model with the embedding \K\, potential \rf{N1} is that at $\xi=0$ the embedding \K\, potential has an enhanced symmetry  \rf{sym2}
 between the inflaton and a conformon. When the local conformal symmetry is spontaneously broken, in supergravity this symmetry is reflected in a shift symmetry of the physical \K\, potential, it depends only on $\Phi-\bar \Phi$. Note, however, that an arbitrary supergravity with the  shift symmetry of the physical \K\, potential is not associated with a particular symmetry  in  in the Jordan frame with non-minimal coupling of gravity.

 {\it Supersymmetry, Jordan/Einstein frame, supergravity and  stabilization of moduli}
 
 To focus on local conformal symmetry we have simplified the model above by postponing the discussion of the role of the sgoldstino and the issue of moduli stabilization. Consistent supersymmetry with ${\cal W}(\Phi, S)= \sqrt \lambda \, \, S\,  \Phi^2\, $ and  $V= \Big |{\partial {\cal W}(S, \Phi)\over \partial {S}}\Big |^2=\lambda (\Phi \bar \Phi)^2$, requires to add a quartic in $S$ term to ${\cal N}$ 
   \be
\mathcal{N}(X,\bar X)= -|X^0|^2 + |\Phi|^2 + |S|^2 - 3 \Delta |X^0|^2  \left[ \left(\frac{\Phi
}{X^{0}}\right)^{2}+\left(\frac{\bar\Phi
}{\bar X^{\bar 0}}\right)^{2}\right] - 3 \zeta {(S\bar S)^2\over |X^0|^2}\,  \,.
\label{calNminimal2}
 \ee
This allows to prove  that $S=0$ is a minimum of the potential.
Note that out of 2 complex scalars, $S$ and $X^1$ only one of them is a light inflaton, the other 3 have to be heavy so that their evolution does not affect inflation and the model is reduced to a single inflaton model.   

Using the condition $X^0=\bar X^{\bar 0}= \sqrt 3 M_P$ we bring the model to a Jordan frame, starting with \rf{calNminimal2}. To get to an Einstein frame from the superconformal model we can use the generic relation \rf{kahler} between $\Omega(z, \bar z)$ and the \K\, potential. In our case
   \be
\Omega(z, \bar z)= -{1\over 3} \mathcal{N}(X,\bar X)|_{X^0=\bar X^{\bar 0}= \sqrt 3 M_P}= e^{-{1\over 3} K(z, \bar z)}\,  \,.
\label{Om}
 \ee
For example, in case of $\Delta>0$ supergravity is defined by
\be
K= -3\log\Big ( 1+ \xi  ( \Phi
^2+\bar{\Phi }^2 
)  -{1\over 3} |S|^2 + {1\over 6} (\Phi - \bar \Phi)^2
+ {\zeta \over 3 }(S\bar S)^2\Big )
\, , \qquad W= \sqrt \lambda \, S \, \Phi^2 \ .
\label{sup}\ee
 If $\zeta\geq 1/6$, the minimum of the potential with $\Phi= {1\over \sqrt 2}(\phi_1+i\phi_2)$   is at  $S=\phi_2=0$, as was established in \cite{Kallosh:2013pby}. In case of $\Delta<0$ it is at  $S=\phi_1=0$. Thus the coupling of a single real field $\phi$ in the
 action \rf{SimplescGravJ2} is recovered from supergravity model defined in  \rf{sup}. This supergravity originates 
  from the superconformal  model \rf{sc}  upon spontaneous breaking of a superconformal symmetry.

\subsubsection {Chaotic inflation in the theory with $V=\lambda^2  f^2 (\phi)$  with non-minimal coupling to gravity ${\xi\over 2} f(\phi)\,   R$, \cite{Kallosh:2013tua}}
It turned out that the model above with a conformal potential $\phi^4$ and non-minimal coupling to gravity ${\xi\over 2}\phi^2\,   R$ is not the only one which has a dramatic improvement in agreement with the data due to $\xi\neq 0$. The class of models
 \begin{align}
 & \mathcal{L}_{\rm J} = \sqrt{-g} [ \tfrac12  \Omega(\phi) R  - \tfrac12 (\partial \phi)^2 -  V_J(\phi) ] \,, \label{Baction} 
\end{align}
with
\begin{align}
& \Omega(\phi)  = 1 + \xi f(\phi)\, ,  \qquad V_J(\phi)=\lambda^2 f^2(\phi) \,.
\end{align}
also shows that  when $\xi $ is increasing,  the agreement with the data on $n_s$ and $ r$  improves significantly.
Here we will present a superconformal version of this model as well as a supergravity one. 
 
The \K\, potential of the embedding space corresponding to a class of models in \cite{Kallosh:2013tua}\footnote{In \cite{Kallosh:2013tua} we introduced a rather useful stabilization term quartic in $S$ associated with $\zeta$, whose advantages are discussed there.} is
   \bea
&&\mathcal{N}(X,\bar X)= -|X^0|^2 + |\Phi|^2 + |S|^2 - {1\over 12}  |X^0|^2  \left[ \phi^{2}+\bar \phi^{2}\right] \nonumber \\
&& - \xi  |X^0|^2  \left[   f (\phi)
+  f(\bar \phi)\right]
- 3 \zeta {(S\bar S)^2\over |X^0|^2 \Big[\Omega (\phi) + \Omega (\bar \phi) \Big]} 
 \,  \,.
\label{general}
 \eea
 where we are using the following notation for the complex field, which has a zero conformal weight under local Weyl transformations
 \be
\phi\equiv \sqrt{6} \,  { \Phi\over X^0} 
 \ee
and the superpotential is
\be
{\cal W}=  \lambda S (X^0)^2 f(\phi)\, .
\ee
Note that the superpotential depends on $X^0$, except in the case that $f\sim \phi^2$, which means a deviation from the CSS.
This model in the Einstein frame leads to a  supergravity with $X^0= \sqrt 3$ and $\phi=\sqrt{2}\,  \Phi $
 \begin{align}
\hskip -10pt K =  & - 3 \log[ \tfrac12 ( \Omega(\phi) + \Omega( \bar \phi))   - \tfrac13 S \bar S + \tfrac12 (\phi - \bar \phi)^2 \notag \\
& +  \zeta \frac{( S \bar S)^2}{\Omega(\phi) + \Omega( \bar \phi)}] \,, \qquad  \qquad 
  W = \lambda S f( \phi)  \,,
\label{KW2} \end{align}
where $\Omega(\phi) = 1 + \xi f(\phi) $ and $f(\phi)$ is a real holomorphic function. This leads exactly to the bosonic model \rf{Baction} discussed in  \cite{Kallosh:2013tua} and in sec. 12.2 in \cite{A} above upon identifying $\Phi = \phi / \sqrt{2}$ while $S=0$, 
which is a consistent truncation as shown in  \cite{Kallosh:2013tua}. 

The superconformal version of this model explains the simplicity of the Jordan frame potential in these models:  in a gauge where the conformon is fixed, the superconformal potential is given by ${\cal W} =  \lambda S  f  (\phi )$   This model generalizes the supersymmetric embedding of the $\phi^4$ theory considered in \cite{Kallosh:2013pby} to arbitrary scalar potentials.

This superconformal/supergravity embedding goes some way towards an understanding of the symmetries underlying the attractor behavior. In particular, for $\xi = 0$ there is symmetry enhancement in the \K\,  potential: it has a shift symmetry in the real part of $\Phi$ and hence does not depend on the inflaton. The same holds for any value of $\xi$ when choosing the function $f(\phi)$ to be a constant. Any deviations from this will introduce a spontaneous breaking of this symmetry. Note also that at $\xi=\zeta=0$ \rf{general} becomes
   \bea
&&\mathcal{N}(X,\bar X)= -|X^0|^2 + |\Phi|^2 + |S|^2 -  3 \Delta |X^0|^2  \left[ \left(\frac{\Phi
}{X^{0}}\right)^{2}+\left(\frac{\bar\Phi
}{\bar X^{\bar 0}}\right)^{2}\right]\label{general1}
 \eea
with $\Delta=\Delta_{cr}= {1\over 6}$. 

\subsection{Deviation from the critical point $\Delta_{cr}=0$, T-models, \cite{Kallosh:2013hoa} }

The bosonic inflationary model has a potential which is an arbitrary function $F(\tanh{\varphi\over \sqrt 6})$. \begin{equation}\label{chaotmodel}
L = \sqrt{-g} \left[  \frac{1}{2}R - \frac{1}{2}\partial_\mu \varphi \partial^{\mu} \varphi -   F(\tanh{\varphi\over \sqrt 6}) \right].
\end{equation}
This model has a property that the values of $n_s, r$ do not depend on the choice of the function $F$ (with some exceptions), in the  approximation where higher orders of $1/N$, where  $N$ is the number of e-foldings, are neglected.
The physical observables in this class of models have an attractor behavior in the large-$N$ limit
 \begin{align}\label{attractorT}
  n_s & \approx 1-\frac{2}{N} \,, \qquad
  r \approx   \frac{12}{N^2} \,.
 \end{align}
 For $N \sim 60$, these predictions $n_s   \sim 0.967$, $r  \sim 0.003$ ($n_s   \sim 0.964$, $r  \sim 0.004$ for $N \sim 55$) are in a perfect agreement with  Planck 2013 data.

Here we describe the superconformal origin of the  bosonic universality class models \cite{Kallosh:2013hoa}. The embedding \K\, potential for these models has an $SU(1,1)$ symmetry between the complex conformon $X^0$ and the complex inflaton $X^1$ superfields (it  is violated only by a quartic in $S$ term, which is vanishing during inflation)
\be
\mathcal{N}(X,\bar X)= -|X^0|^2 + |X^1|^2 + |S|^2  - 3 \zeta\,  {(S\bar S)^2\over |X^0|^2-|X^1|^2 }\,  \,.
\label{calNT}
 \ee
 Thus, the embedding \K\, potential is flat, apart from the $\zeta$-term. It means that, comparing it for example with  \rf{general1}, it corresponds to $\Delta=0$. It is a critical point in the following sense: at $\Delta=0$ there is an enhancement symmetry, the  embedding \K\, potential has an $SU(1,1)$ symmetry (at $\zeta=0$), which is absent at arbitrary non-vanishing $\Delta $.

We take the superpotential   which preserves a subgroup of $SU(1,1)$, which is $SO(1,1)$. For $f=$const    the boost  between the holomorphic parts of $X^0$ and $X^1$ is preserved.  
\be
{\cal W} = S \Big ((X^0)^2- (X^1)^2\Big ) f (X^1/X^0)  \ .
 \label{super}\ee
The superpotential depends on $X^0$, which is a deviation from the CSS.
 Function  $f (X^1/X^0)$ is invariant under local conformal-$\mathbb{R}$-symmetry, but when it is not a constant, it deformes the $SO(1,1)$.

\subsubsection{Conformal-$\mathbb{R}$-symmetry breaking condition $X^0= \bar X^0= \sqrt 3$}

We  spontaneously break local conformal as well as a local $U(1)$ symmetry by taking $X^0= \bar X^0= \sqrt 3$ condition and with $X^1\equiv \Phi$ we recover, according to \rf{kahler}, a supergravity version of the superconformal model with 
\be
K= -3 \ln \Big [ 1- {|\Phi |^2 + |S^2|\over 3} + \zeta \, {(S\bar S)^2\over 3-|\Phi|^2 } \Big],
\label{K}\ee
and 
 \be
 W = S \Big (3- (\Phi )^2\Big ) f (\Phi/\sqrt 3) \ .
 \ee
An advantage of this choice is that it is easy to  study the moduli stabilization since in standard supergravity the relevant codes were used for a long time. In this case the inflaton is a real part of $\Phi$.
In particular at $S=0$ our condition that $1- {|\Phi |^2 |\over 3}>0$ is a condition that we are inside a `\K\, cone'. The kinetic term for the scalar $\Phi$ is 
\be
\Big({1\over 1-|\Phi |^2 / 3}\Big )^2 \partial_\mu \Phi \partial^\mu \bar \Phi \ .
\ee 
The positivity of the kinetic term for scalars requires that $|\Phi |^2<3$. The boundary of the moduli space here is a `\K\, cone'
\be
1- {|\Phi |^2 |\over 3}=0 \ .
\ee
One finds that the condition for stability of a sgoldstino at $S=0$ for all values of the inflaton  is provided by $\zeta>1/6$ in \rf{K}. Inflation is stable at $\rm Im \, \Phi=0$ independently of the value of $\zeta$.  

\subsubsection{   $SO(1,1)$-invariant symmetry breaking  rapidity condition $(X^0)^2 - (X^1)^2 = 3$}

Now we use the fact which we learned above:  inflation takes place at
\be
S=0\, ,  \qquad X^1= \bar X^1= \varphi/\sqrt 2 \ .
\ee
We may resolve the $SO(1.1)$-invariant  constraint $(X^0)^2 - (X^1)^2 = 3$  (rapidity condition) so that 
\be
X^0= \sqrt 3 \cosh   \varphi/ \sqrt 6 \ , \qquad X^1= \sqrt 3 \sinh   \varphi/ \sqrt 6 \ .
\ee
The superconformal action \rf{sc}
with \rf{calNminimal2} and \rf{sup} entries becomes at 
$
S=0 $,  $X^1= \bar X^1= \varphi/\sqrt 2
$
\be
{1\over \sqrt{-g}}\mathcal{L}_{\rm sc}^{\rm scalar-grav}={1\over 2}
R
-{1\over 2} (\partial^\mu \varphi)^2 -  |f (\tanh(\varphi/ \sqrt 6))|^2 \, .
 \label{scGrav11}
\end{equation}
This provides a very simple relation between the superconformal theory and the bosonic universality class models \rf{chaotmodel} in case that $F= f\bar f $.

We will call these models $\alpha=1$ attractors, when comparing them to the class of model below.

\subsection{Superconformal $\alpha$-attractors \cite{Kallosh:2013yoa}}

We now turn to the generalization of these $\tanh(\varphi/ \sqrt 6)$ superconformal models, leading to a family of $\alpha$-attractors. The deformation of the CSS belongs to the more general class 3, as explained in the beginning of sec. 3. At $\alpha=1$ we recover T-models presented above.

The superconformal \K\,  potential of the embedding space is now given by
 \begin{align}
  \mathcal{N}(X, \bar X)  = - |X^0|^2 \left[ 1 - \frac{|X^1|^2 + |S|^2}{|X^0|^2} \right]^\alpha \,.
 \end{align}
Note that the \K\, potential only preserves the manifest $SU(1,1)$ symmetry between $X^0$ and $X^1$  for the special value $\alpha = 1$. The superconformal superpotential reads
 \begin{align}
  \mathcal{W} = S (X^0)^2 f(X^1 / X^0) \left[ 1 - \frac{(X^1)^2}{(X^0)^2} \right]^{(3 \alpha -1)/2} \,.
\label{calW} \end{align}
 The superpotential with a constant $f$ and $\alpha =1$ preserves the $SO(1,1)$ symmetry, the subgroup of $SU(1,1)$. However, when either $f$ is not constant, or $\alpha \neq 1$,  the $SO(1,1)$ symmetry is deformed.

In order to extract a Poincar{\'e} supergravity we impose the  $X^0 = {\bar X}^0 = \sqrt{3}$ condition. The \K\, and superpotential are then given by 
 \begin{align}
  K & = - 3 \alpha \log \left[1 - \frac{S \bar S+ \Phi \bar \Phi}{3} \right] \,, \qquad
  W = S f(\Phi / \sqrt{3}) (3 - \Phi^2)^{(3 \alpha -1)/2} \,.
\label{KW1} \end{align}
For a generic real functions $f$, the   model above allows  a truncation to a one-field model via $S = \Phi - \bar \Phi = 0$; the stability of this truncation will be discussed below.
The  effective Lagrangian at $S=\Phi - \bar \Phi=0$ is
 \begin{align}
  \mathcal{L} = \sqrt{-g} \left[ {1\over 2} R - \frac{\alpha}{ (1- \Phi^2 / 3)^2}(\partial \Phi)^2 - f^2(\Phi / \sqrt{3})  \right] \, \,.
 \end{align}
Therefore the action is greatly simplified for real $\Phi$. As in  \cite{Kallosh:2013hoa}  we find a simple relation between the geometric field $\Phi$ and a canonical one $\varphi$: it is the rapidity-like relation 
\be
  {\Phi \over \sqrt{3}} = \tanh {\varphi \over \sqrt {6 \alpha}} \,.
\label{rapidity}
\ee
This is fully analogous to the relation between velocity $v$ and rapidity $\theta$ in special relativity, ${v\over c} =  \tanh \theta$. The geometric non-canonical field ${|\Phi|} < \sqrt{3}$ has a limited range, analogous to velocity bound $v< c$;  in contrast, the rapidity and the canonical field $\vp$ have an unlimited range. 
The action for a canonical field  $ \varphi $ has an effective Lagrangian
 \begin{align}
  \mathcal{L} = \sqrt{-g} \left[ {1\over 2} R - {1\over 2}  (\partial \varphi)^2 - f^2\big(\tanh {\varphi\over\sqrt{6\alpha}}\big)
 \right] \, . 
 \label{action}  \end{align} 
At $f=$ const the potential of this model does not depend on $\vp$ and $\alpha$ and describes de Sitter vacuum.

In order to understand the role of the $\alpha$ parameter, we note that all $\alpha$-models during inflation at $S=0$ are defined by the $SU(1,1) / U(1)$  \K\, potential of the inflaton multiplet
\begin{align}
  K & = - 3 \alpha \log \left(1 - \frac{\Phi \bar \Phi}{3} \right) \,.   \end{align}
 This leads to kinetic terms of the form
\be
K_{\Phi\bar{\Phi}}\partial \Phi \partial \bar \Phi={\alpha\over  \bigl(1 - \frac{\Phi \bar \Phi}{3} \bigr)^2}  \partial \Phi \partial \bar \Phi \ .
\ee
This \K\, metric $g_{\Phi\bar \Phi}= K_{\Phi\bar{\Phi}}$ corresponds to an $SU(1,1)/U(1)$ symmetric space with the constant curvature:
 \be
R_{K}= -g^{-1}_{\Phi\bar \Phi}\partial_\Phi\partial_{\bar \Phi} \log g_{\Phi\bar \Phi}= - {2\over 3 \alpha}  \,,
 \ee
The same relation was found in the context of the supersymmetric $\alpha$-$\beta$ model in \cite{Ferrara:2013rsa}. Here we notice that
 all  $\alpha$-attractors  with an arbitrary function $f\big (\tanh{\varphi\over \sqrt{6\alpha}}\big)$
 have a universal interpretation of the parameter $\alpha$,
 \be
\alpha =  -{2\over 3R_{K}}  \,,
 \ee
in terms of the $SU(1,1)/U(1)$ symmetric space with the negative constant curvature $R_K$ in this class of models.

Finally, we address the stability of the truncation to the single-field model. To this end we add a stabilisation term to our original superconformal \K\, potential,
 \begin{align}
  \mathcal{N}(X, \bar X)  = - |X^0|^2 \left[ 1 - \frac{|X^1|^2 + |S|^2}{|X^0|^2} + 3 g\frac{|S|^4}{|X^0|^2 ( |X^0|^2 - |X^1|^2)} \right]^\alpha \,.
 \end{align}
The original four scalar fields have the following masses at the inflationary trajectory $S = \Phi - \bar \Phi = 0$:
 \begin{align}
  m_{{\rm Re}(\Phi)}^2 & = \eta_\varphi V \,, \quad
  m_{{\rm Im}(\Phi)}^2  = \left(2 - \frac{2}{3 \alpha} + 2 \epsilon_\varphi - \eta_\varphi \right) V \,, \quad
  m_S^2 = \left( \frac{12 g -2}{3 \alpha} + \epsilon_\varphi \right) V \,,
 \end{align}
where $\epsilon_\vp$ and $\eta_\vp$ are the slow-roll parameters of the effective single-field model \eqref{action}. In order to achieve stability up to slow-roll suppressed corrections, the second equation requires $\alpha > 1/3$ for stabilisation of the inflationary trajectory, and the latter requires $g > 1/6$. 

The physical observables in this class of models have an attractor behavior in the large-$N$ limit, they do not depend on the generic choice of the function $f\big(\tanh {\varphi\over\sqrt{6\alpha}}\big)$
 \begin{align}\label{attractor}
  n_s & \approx 1-\frac{2}{N} \,, \qquad
  r \approx  \alpha \frac{12}{N^2} \,.
 \end{align}
 These expressions are valid for $\alpha$ not significantly  greater  than 1.
 However, the level of gravity waves depends linearly (at large $N$) on the inverse curvature of the \K\, manifold.
Note that  the supersymmetric models \rf{KW1} are stabilized at  $\alpha>1/3$. It is interesting to look at  the range of the attractor points near $\alpha=1$ where the value of $r$ changes by the order of magnitude:
\be
1/3 <\alpha <3 \,, \qquad         10^{-3} < r < 10^{-2} \ .
\ee
The class of these models  represents a  generalization of the attractor values \eqref{attractorT}, which have appeared in a variety of contexts, to the family of attractor values \rf{attractor} labelled by the parameter $\alpha$. The so-called  $\alpha - \beta$ models in  \cite{Ferrara:2013rsa} also belong to this class, although the superconformal and supergravity embedding is different.

This models have the same behavior for sufficiently small $\alpha$,  for example for $(\tanh {\vp \over \sqrt {6\alpha}})^n $ the observables do not depend on $n$. The potential in  the new class of $\alpha$-attractors  is {\it an arbitrary function of $\tanh {\vp \over \sqrt {6\alpha}}$}.
For large $\alpha$ these models yield conventional chaotic inflation. For example for $(\tanh {\vp \over \sqrt {6\alpha}})^n $ at large $\alpha$  the argument of $\tanh$ becomes very small so that 
$(\tanh {\vp \over \sqrt {6\alpha}})^n \sim \vp^n$.
These models provide a continuous interpolation between the chaotic inflationary values for $(n_s,r)$ and the universal attractor values \eqref{attractor}. Moreover, the parameter $\alpha$ has the  geometrical interpretation, corresponding to the inverse curvature of the underlying scalar manifold of the inflaton's supermultiplet. Its role in the present discussion is to control the distinction \eqref{rapidity} between the geometric and canonical fields, akin to that between velocity and rapidity in special relativity.

 Therefore for a large set of choices for $f^2(\tanh {\vp \over \sqrt {6\alpha}})$ with sufficiently small $\alpha$, we find the universal  attractor in the $r$-$n_s$ plane \eqref{attractor}. A number of regimes is of particular interest:
 \begin{itemize}
\item
For $\alpha=1$ and  $R_{K}=  -{2\over 3  }$ we have a special case when the underlying \K\, potential of the embedding manifold has an unbroken $SU(1,1)$ symmetry and the superpotential has an $SO(1,1)$ subgroup of this symmetry deformed only by the deviation of the function $f(\tanh {\vp \over \sqrt {6}})$ from a constant value. 
\item 
For moderate deviations from this special value, of the order $1/3 < \alpha < 3$, the level of gravity waves varies universally between $10^{-3} < r < 10^{-2}$. This entire regime, with $-2 < R_K < - {2 \over 9}$, is therefore subject to the attractor behavior.
\item
 Allowing $\alpha$ to grow significantly beyond this upper limit, the attractor behavior is lost and the values for $(n_s,r)$ start depending on the choice of the function $f(\tanh {\vp \over \sqrt {6\alpha}})$. In this case the curvature of the \K\, manifold decreases.
 \item
 If we instead take $\alpha$ smaller than $1/3$, in the context of the present superconformal realization of the $\alpha$-attractors with chiral superfields, instead of the vector one, the scalar superpartner of the inflaton has a negative mass squared: the model becomes unstable in this limit with $\alpha < 1/3$. However,
 the bosonic model, as well as the $\alpha - \beta$ model \cite{Ferrara:2013rsa},  still display the attractor behaviour with the universal values, which approach $r=0$ as $\alpha \rightarrow 0$. In this limit the curvature of the \K\,  manifold in $\alpha - \beta$ model \cite{Ferrara:2013rsa}, $R_{K}$,  increases. 
 \end{itemize}
Both from the current as well as other perspectives it would therefore be of utmost interest to eventually learn the value of $r$ in the case of a detection of gravity waves, or the value of improved upper bounds on $r$. In the context of the current class of models, this would constitute a constraint on the curvature of the \K\, manifold $R_K$.  Meanwhile, any improvement in the error margin for the value of $n_s$ can be used to determine the number of e-folds $N$ more accurately.

\section{Discussion}

The new data release by Planck 2013  suggests that a single field inflation is a plausible candidate for explaining the cosmological observations.
The simplicity of the early universe might be caused by some hidden symmetry, which would explain why the universe is `almost perfect'. In this lecture we were trying to explain why the superconformal symmetry underlying the four-dimensional supergravity is a useful concept which makes the non-minimal coupling of scalars to gravity a natural part of construction of inflationary models. It remains to be seen whether it is a fundamental or just a very useful concept. 

Superconformal symmetry appears to be useful for the classification of various models of supergravity which under certain conditions become cosmological attractors. This symmetry certainly supports the concept of the non-minimal coupling of scalars to curvature, as we explained in this lecture. Non-minimal coupling may substantially reduce the level of predicted tensor modes for all models of chaotic inflation.
Having at our disposal the flexibility in the choice of the non-minimal coupling to gravity $\xi$ will allow an easy interpretation of many inflationary models if the gravity waves are detected at  $r\ll10^{-1}$. 

The superconformal symmetry leads to the discovery of the cosmological attractors, numerous models with the same predictions for the observables $n_{s}$ and $r$.
For example, the $\alpha$-attractors have the following  attractor values for the observables 
 \begin{align}\label{attractor1}
  n_s & \approx 1-\frac{2}{N} \,, \qquad
  r \approx  \alpha \frac{12}{N^2} \,.
 \end{align}
where $N$ is the number of e-foldings of inflation and $\alpha$ is a parameter related to the curvature of the \K\, manifold.

The fact that these predictions are not for a single model of inflation but for a large class of them, for example for  a large set of choices of functions $f^2(\tanh {\vp \over \sqrt {6\alpha}})$, makes it more desirable to test the region for detection of primordial gravity waves with $10^{-3} \lesssim r   \lesssim 10^{-1}$. 

Stringy models of inflation were described in the recent review \cite{Burgess:2013sla}  and in the lectures in this volume by E. Silverstein \cite{Silverstein:2013wua}. These models are expected to have a UV completion. In majority of the string inflation models the prediction for gravity waves is well below $r\sim 10^{-5}$, which makes the detection impossible. Meanwhile, there are few very interesting string inflation models, based on monodromy concept \cite{Silverstein:2008sg}, which predict $r\sim 7\cdot 10^{-2}$ and $3 \cdot 10^{-2}$. These models are likely to be tested experimentally in the near future. Below that, we have a full region  of our new cosmological attractors, where there are many models predicting $r$ below the monodromy level models, and still detectable, maybe during the next decade or so. 
The prediction of the Starobinsky model \cite{Starobinsky:1980te} with $r\sim 4\cdot 10^{-3}$ gave a sufficient reason to look for this level of detection. The fact that chaotic inflation in the theory $\lambda\phi^{4}$ with nonminimal coupling to gravity with $\xi \gtrsim 0.1$ leads to the same prediction gave us an additional reason to look in this direction.  The discovery of the large family of cosmological attractors which lead to identical or very similar predictions  makes the goal of detecting gravity waves from inflation in the range of $10^{-3} < r < 10^{-2}$ much more interesting and promising than before.

\subsubsection*{Acknowledgments}
 
I am grateful to the organizers of the Les Houches School ``Post-Planck Cosmology'' in July-August 2013 for 
their hospitality. I would like to thank  R. Bond, G. Efstathiou, S. Mukhanov and  E. Silverstein  for most useful discussions and S. Ferrara, A. Linde, D. Roest and A. Van Proeyen for collaboration on work presented in this lecture.  This work was supported by the NSF Grant PHY-1316699, SITP and Templeton grant ``Quantum Gravity frontier''.

\end{document}